\documentclass{emulateapj}
\usepackage{natbib}

\usepackage{epsfig}
\usepackage{rotating}

\newcommand{\scl}	{\Sigma_{\rm cl}}
\newcommand{\gcm} {{\rm g\:cm}^{-2}}
\newcommand{\myr} {M_\odot\:\mathrm{yr}^{-1}}
\newcommand{\myrkms} {M_\odot {\rm yr^{-1} km\:s^{-1}}}

\begin{document}

\title{A Massive Protostar Forming by Ordered Collapse of a Dense, Massive Core}

\author{Yichen Zhang$^1$, Jonathan C. Tan$^{1,2}$, James M. De Buizer$^3$, G\"oran Sandell$^3$,
	Maria T. Beltran$^4$, Ed Churchwell$^5$, Christopher F. McKee$^{6,7}$, 
	Ralph Shuping$^3$, Jan E. Staff$^8$, Charles Telesco$^1$, Barbara Whitney$^5$}
\affil{$^1$Department of Astronomy, University of Florida, Gainesville, Florida 32611, USA\\
$^2$Department of Physics, University of Florida, Gainesville, Florida 32611, USA\\
$^3$SOFIA-USRA, NASA Ames Research Center, MS 232-12, Building N232, P. O. Box 1, 
Moffett Field, CA 94035, USA\\
$^4$INAF-Osservatorio Astrofisico di Arcetri, Largo E. Fermi 5, Firenze I-50125, Italy\\
$^5$Department of Astronomy, University of Wisconsin, Madison, Wisconsin 53706, USA\\
$^6$Department of Astronomy and Astrophysics, University of California, Berkeley, California 94720, USA\\
$^7$Department of Physics, University of California, Berkeley, California 94720, USA\\
$^8$Department of Physics and Astronomy, Louisiana State University, Baton Rouge, Louisiana 70803, USA\\
yc.zhang@astro.ufl.edu}

\begin{abstract}
We present 30 and 40~$\mu$m imaging of the massive protostar
G35.20-0.74 with SOFIA-FORCAST. The high surface density of the natal
core around the protostar leads to high extinction, even at these
relatively long wavelengths, causing the observed flux to be dominated
by that emerging from the near-facing outflow cavity. However,
emission from the far-facing cavity is still clearly detected. We
combine these results with fluxes from the near-infrared to mm to
construct a spectral energy distribution (SED). For isotropic emission
the bolometric luminosity would be $3.3\:\times\:10^4\:L_\odot$.  We
perform radiative transfer modeling of a protostar forming by ordered,
symmetric collapse from a massive core bounded by a clump with high
mass surface density, $\scl$. To fit the SED requires
protostellar masses $\sim\:20-34\:M_\odot$ depending on the outflow
cavity opening angle ($35^\circ-50^\circ$), and $\scl\sim0.4-1\:\gcm$. 
After accounting for the foreground extinction and the
flashlight effect, the true bolometric luminosity is $\sim(0.7-2.2)\times\:10^5\:L_\odot$. 
One of these models also has
excellent agreement with the observed intensity profiles along the
outflow axis at 10, 18, 31 and 37~$\mu$m. 
Overall our results support a model of massive star formation
involving the relatively ordered, symmetric collapse of a massive,
dense core and the launching bipolar outflows that clear low density
cavities. Thus a unified model may apply for the formation of both low
and high mass stars.
\end{abstract}


\section{Introduction}\label{S:intro}

Massive stars impact many areas of astrophysics. In most galactic
environments they dominate the radiative, mechanical and chemical
feedback on the interstellar medium, thus regulating the evolution of
galaxies. Many low-mass stars form in clusters along with massive stars, and
their protoplanetary disks can be affected by this feedback
also. There is some evidence that our own solar system was influenced
in this way (\citealt[]{Adams10}). Despite this importance, there is
no consensus on the basic formation mechanism of massive
stars. Theories range from {\it Core Accretion}, i.e., a scaled-up
version of low-mass star formation (e.g. the Turbulent Core Model of
\citealt[MT03]{MT03}), to {\it Competitive Accretion} at
the centers of forming star clusters (\citealt[]{Bonnell01},
\citealt[]{Wang10}), to {\it Stellar Collisions}
(\citealt[]{Bonnell98}).  

The Core Accretion theory predicts the existence of an envelope-fed,
central accretion disk and relatively ordered and collimated bipolar
outflows powered by accretion around a massive protostar. The other
formation mechanisms predict a much less ordered geometry of gas and
dust surrounding the protostar.

Collimated bipolar outflows have been observed from massive protostars
(e.g. \citealt[]{Beuther02}). Outflows may limit the formation
efficiency (mass ratio of final star to initial gravitationally-bounded 
protostellar core) to $\sim$0.5 (\citealt[]{MM99}), since they
expel material along polar directions. The resulting low-density
cavities have been proposed to significantly affect the appearance of
the protostar in the MIR (\citealt[]{Debuizer06}, \citealt[]{Debuizer12}), and this
is seen in the radiative transfer (RT) calculations of \citet[ZT11]{ZT11} and
\citet[ZTM13]{ZTM13} using the RT code developed by \citet[]{Whitney03}.

The Turbulent Core Model also relates core properties to the
surrounding self-gravitating star-cluster-forming
clump.  For a marginally unstable pre-stellar core of a given mass,
its size, density and subsequent accretion rate all depend on the
pressure set by the mean mass surface density of the clump, $\scl$. 
Observed values of $\scl$ are $\sim0.1-1\:\gcm$
 (MT03; \citealt[]{BT12}), leading to high
predicted accretion rates, $\sim10^{-4}-10^{-3}\:\myr$, 
potentially important for overcoming the high radiation
pressure from the massive protostar. 

G35.20-0.74 (hereafter G35.2) is a massive protostar associated with a
well-defined outflow. At a distance of 2.2~kpc (\citealt[]{Zhang09}),
radio continuum emission shows that it contains an ultracompact HII
region at the center in the form of a collimated jet along the
north-south direction (\citealt[]{Heaton88}, \citealt[]{Gibb03}),
which may be an example of an {\it Outflow-Confined HII Region}
(\citealt[]{TM03}).  The outflow is also observed in the NIR
(\citealt[]{Fuller01}) with the northern cavity brighter than the
southern one. Ground-based 10 and 18~$\mu$m observations only reveal
the northern cavity (\citealt[]{Debuizer06}), suggesting it is
inclined towards us. Outflow is also seen in CO (\citealt[]{Gibb03},
\citealt[]{Birks06}), including a wider angle flow in the NE-SW
direction. It is unclear if this results from a separate driving
source or is related to the N-S jet, but misaligned by interaction
with the surrounding core/clump. In previous studies, an early B star
was thought to be present at the center (\citealt[]{Dent85}),
surrounded by an envelope with mass $\sim\:500\:M_\odot$
(\citealt[]{Little98}) to $\sim\:3800\:M_\odot$ (\citealt[]{Paron10}).

Here we present the first SOFIA-FORCAST images of G35.2. Combined with
other multiwavelength data, we construct the SED. We present a simple
exploration of parameter space of RT models of massive protostars and
use the observed SED and intensity profiles along the outflow axis to
constrain the models.

\section{Observations \& Data Reduction}\label{S:obs}

The SOFIA-FORCAST (\citealt[]{Herter12}) observations of G35.2 were
performed on May 24 and 26 (UT) 2011 in the 11.3, 19.7, 31.5, and
37.1~$\mu$m filters at an altitude of 43000~ft. The chopping secondary
of SOFIA was driven at 4~Hz, with a matched chop and nod throw of
60$\arcsec$, and with a nod performed every 30s. The final effective
on-source exposure times were 909~s at 11.3~$\mu$m, 959~s at
19.7~$\mu$m, 4068~s at 31.5~$\mu$m, and 4801~s at 37.1~$\mu$m.  The
fluxes were calibrated by the SOFIA data reduction pipeline, which
during the Basic Science period has an estimated 3$\sigma\leq20\%$
calibration error in all filters (\citealt[]{Herter12}).

\begin{figure*}
\begin{center}
\includegraphics[width=\textwidth]{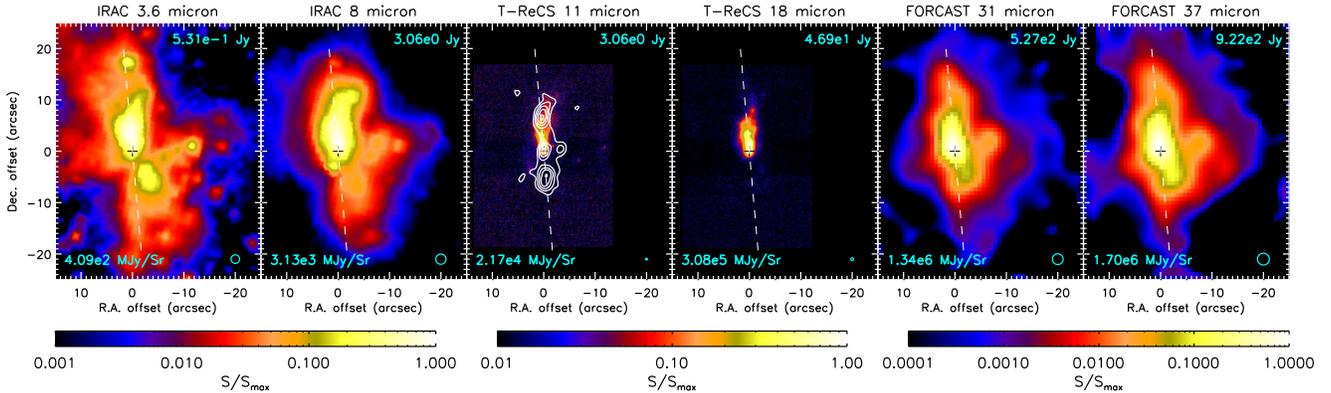}\\
\caption{Multiwavelength images of G35.2. From left to right,
IRAC (GLIMPSE) at 3.6 and 8~$\mu$m;
T-Recs at 11 and 18~$\mu$m (\citealt[]{Debuizer06});
(deconvolved) FORCAST at 31 and 37~$\mu$m.
The white contours in the 11~$\mu$m image show 15 GHz radio
continuum emission (\citealt[]{Heaton88}). Total fluxes are labeled
in the upper-right corners and maximum surface brightnesses in
the lower-left.  The dashed line indicates the axis of the
outflow deduced from the MIR and radio emission. 
Circles indicate image resolutions.}
\label{fig:img}
\end{center}
\end{figure*}

The 11 and 31~$\mu$m images were observed simultaneously with the
dichroic mode of FORCAST; and similarly for the 19 and 37~$\mu$m
images. The relative offsets between the short and long wavelength
arrays are known to better than $1\sigma=0.15\arcsec$.  The best
registration of the 19~$\mu$m image with respect to the 18.3~$\mu$m
image from \citet[]{Debuizer06} was obtained using a $\chi^2$
minimization algorithm.  The latter was convolved with a Gaussian
profile to reach the same resolution of the FORCAST image.  Several
features in the two images match up well after the registration
increasing confidence in the technique. We adopt the pixel size
(0.768$\arcsec$) as our conservative 1$\sigma$ uncertainty in the
relative offset between the 18.3 and 19.7~$\mu$m images.  Considering
the absolute astrometric uncertainties of $\sim0.2\arcsec$ from the
18.3~$\mu$m image, we get an absolute astrometric uncertainty of
0.79$\arcsec$ from the 19.7~$\mu$m image.  The absolute astrometric
uncertainty of the 37.1~$\mu$m image is then estimated to be better
than 0.8$\arcsec$.  However, the 11.3~$\mu$m FORCAST image suffered
from such low S/N that it could not be used to perform similar
registration for the 31.4~$\mu$m image.  We had to use the $\chi$$^2$
minimization algorithm to find the best registration of the
31.4~$\mu$m image and the astrometrically calibrated 37.1~$\mu$m
image, noticing that they look remarkably similar morphologically. The
absolute astrometric uncertainty in the 31.4~$\mu$m image is estimated
to be 0.89$\arcsec$.

\begin{figure}
\begin{center}
\includegraphics[width=\columnwidth]{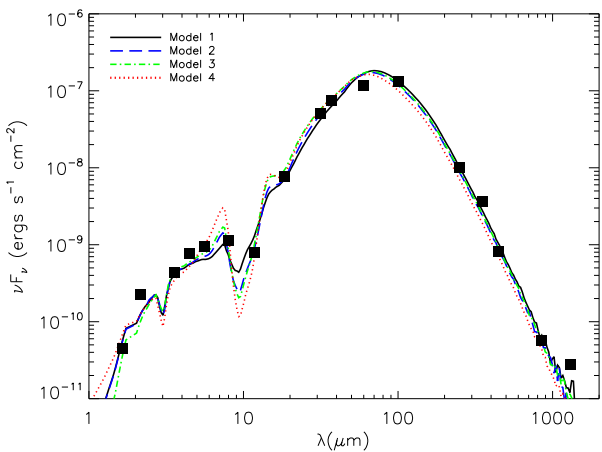}\\
\caption{SED of G35.2 (solid squares) with observed
fluxes from 2MASS H \& K bands, IRAC from 3.5 to 8~$\mu$m, 
T-ReCs 11, 18~$\mu$m, FORCAST 31,
37~$\mu$m, IRAS 60, 100~$\mu$m, SPIRE 250, 350~$\mu$m, SCUBA 450,
850~$\mu$m, and IRAM 1.3~mm. The lines show four model SEDs
(see text).}
\label{fig:sed}
\end{center}
\end{figure}

\begin{figure*}
\begin{center}
\includegraphics[width=\textwidth]{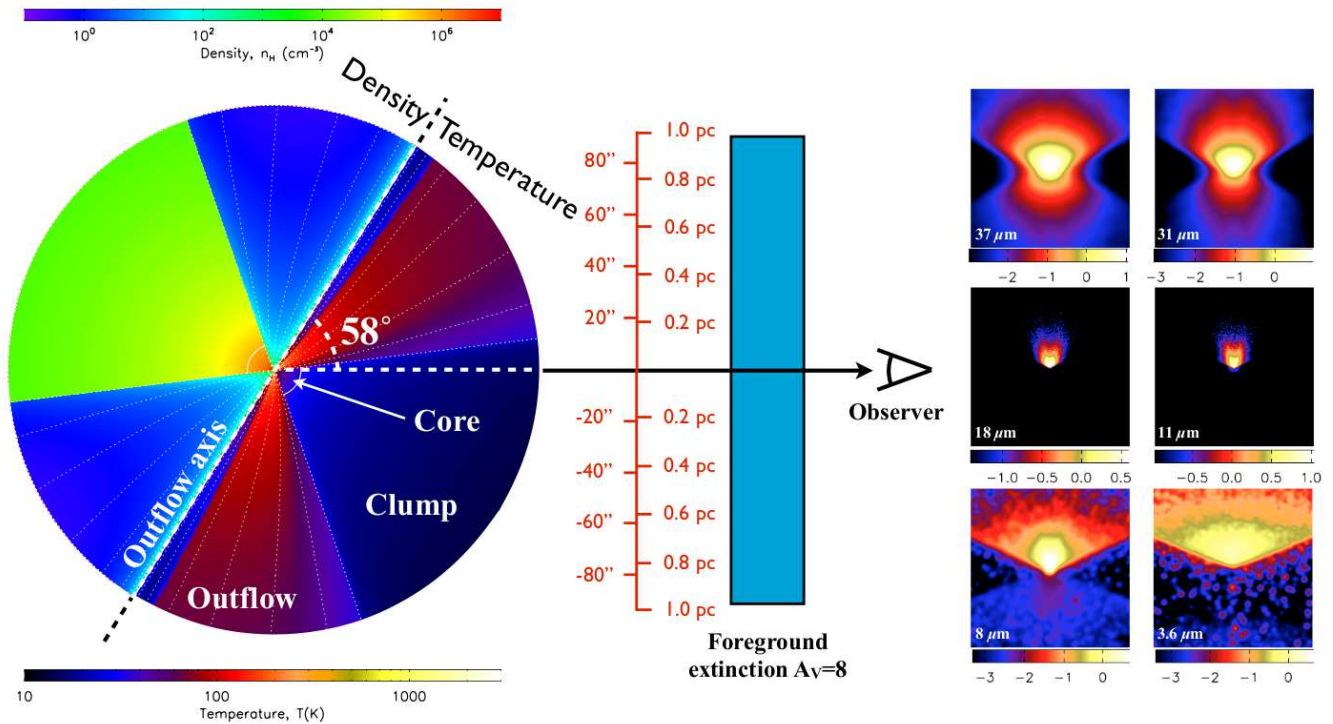}\\
\caption{Cartoon showing geometry of Model 1.
Density and temperature profiles are on the upper-left 
and lower-right sides of the outflow axis, respectively.
White dotted lines show disk wind streamlines.
The low temperature region close to the outflow axis is dust-free wind 
launched from the inner hot dust-free disk. The clump and the outflow extend to 
10 core radii. 
The simulated images ($60\arcsec\times60\arcsec$) of Model 1 
are also shown. 
The scalebars are in $\log(S/S_\mathrm{avg})$,
where $S_\mathrm{avg}$ is the 
average intensities of the near-facing outflow in each band (see Section \ref{S:fluxdist}). 
For the T-ReCs and FORCAST bands,
we apply same dynamic ranges as in Fig. (\ref{fig:img}), 
while for the IRAC bands, we adopt larger dynamic ranges
to show the far-facing outflow.}
\label{fig:cartoon}
\end{center}
\end{figure*}

\begin{figure}
\begin{center}
\includegraphics[width=\columnwidth]{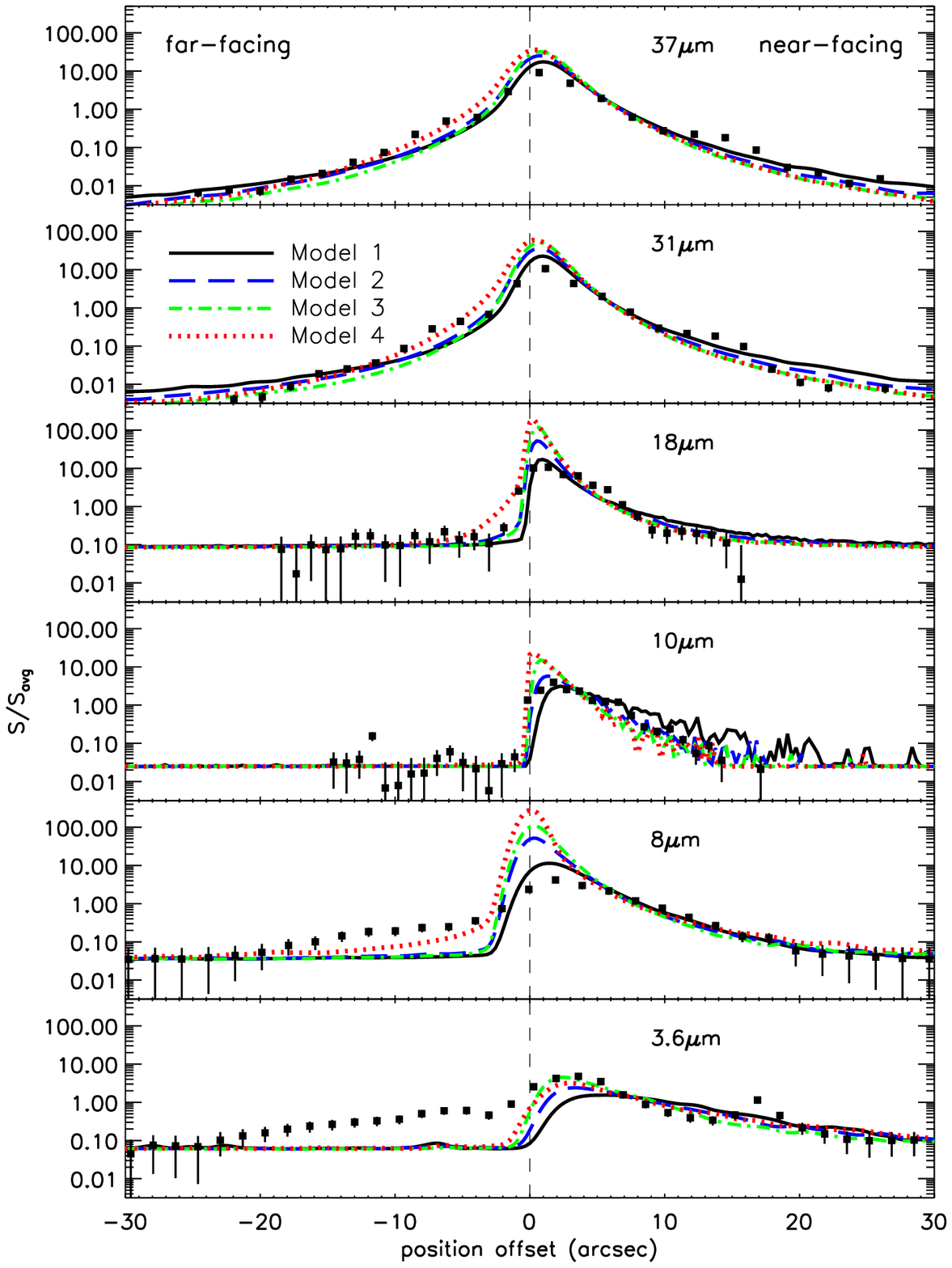}\\
\caption{Intensity profiles along the outflow axis. 
The squares are observational data sampled at intervals
of the resolutions of the instruments (intervals of 2 \& 3 $\times$
resolution are used for 18 \& 10~$\mu$m) with errors composed of
systematic flux uncertainties (assumed to be 20\%) and estimated
background noise. The lines are model profiles.}
\label{fig:fluxdist}
\end{center}
\end{figure}

Using standard stars observed throughout several Basic Science
flights, an average FWHM for each wavelength was determined:
3.9$\arcsec$ at 19.7~$\mu$m, 4.1$\arcsec$ at 31.5~$\mu$m, and
4.5$\arcsec$ at 37.1~$\mu$m.  The images were then deconvolved using
the maximum likelihood method (\citealt[]{Richardson72},
\citealt[]{Lucy74}), which yielded an approximately factor of two
better resolution than the pre-deconvolved images.  The deconvolved
images compare favorably to simple unsharp masking of the original
images, and in the case of the 19.7~$\mu$m image, it compares well to
the Gemini 18.3~$\mu$m image, and thus the substructures revealed in
the deconvolved images are believed to be real.

\begin{table*}
\begin{center}
\caption{Parameters of the four fitted models.}
\label{tab:models}
\begin{tabular}{c|c|c|c|c|c|c|c}
\hline
Model & Core mass & Mean surface density of &
Protostellar & Outflow opening & Inclination & 
Foreground & Luminosity \\
& $M_c$ ($M_\odot$) & the clump $\scl$ ($\gcm$) &
mass, $m_*$ ($M_\odot$) & angle $\theta_{w,\mathrm{esc}}$ & $\theta_\mathrm{view}$
\footnote{Inclination between the line of sight and the outflow axis.} & 
extinction $A_V$ & $L_\mathrm{bol}$ ($L_\odot$)
\footnote{True bolometric luminosity after allowing for 
foreground extinction and flashlight effect.}\\
\hline
Model 1 & 240 & 1 & 34 & 51$^\circ$ & 58$^\circ$ & 8 & $2.2\times10^5$\\
\hline
Model 2 & 240 & 0.7 & 26 & 45$^\circ$ & 51$^\circ$ & 8 & $1.2\times10^5$\\
\hline
Model 3 & 240 & 0.7 & 22 & 40$^\circ$ & 44$^\circ$ & 15 & $9.0\times10^4$\\
\hline
Model 4 & 240 & 0.4 & 20 & 35$^\circ$ & 43$^\circ$ & 0 & $6.6\times10^4$\\
\hline
\end{tabular}
\end{center}
\end{table*}

We also retrieved calibrated level 2 data (AOR 1342241157, HOBYS GT program) 
from Herschel-SPIRE and a single long integration SCUBA 850/450 $\mu$m 
jiggle-map form the JCMT archive at CADC\footnote{Guest user, Canadian Astronomy Data Center}.
The SCUBA data were reduced and calibrated in a 
standard way using SURF (\citealt[]{Sandell01}).

\section{Results}\label{S:results}

Fig.~(\ref{fig:img}) shows the FORCAST images at 31.5 and 37.1~$\mu$m
along with the 11 and 18~$\mu$m images by Gemini-T-ReCS
(\citealt[]{Debuizer06}) and 3.6 and 8~$\mu$m images by Spitzer-IRAC
(GLIMPSE Survey, \citealt[]{Churchwell09}). The airborne FORCAST
images have a dynamic range of $\sim 4$ orders of magnitude, similar
to the space-based IRAC observations and much higher than the
ground-based T-ReCS observations. The origin of the images is chosen
to be the radio source G35.2N (R.A.=18h58m13.033s, DEC.=+1d40m36.14s).
While the 11 and 18~$\mu$m images only show the northern side of the
outflow, the FORCAST 31 and 37~$\mu$m images also reveal the fainter
southern outflow cavity. This cavity is also seen in the IRAC images,
again being fainter than the northern side.

This trend agrees with the results of the RT simulations by ZT11 and
ZTM13, which showed that at $\sim$40~$\mu$m, the near-facing outflow
cavity is the dominant feature but, because of lower extinction, the
structure along the outflow axis is more symmetric than at shorter
wavelengths. In these models, the far-facing outflow appears in the
NIR mainly due to the scattering by dust in and around the outflow
cavity.

With their high dynamic range, the FORCAST images also show extended
emission on larger scales, which is missing in the ground-base MIR
images. The morphology of this extended emission has a resemblance to
that seen with IRAC. This extended emission seems to be related to the
cavity wall and/or a wider-angle part of the outflow. 
It is also possible that some, probably
small, fraction of this emission is produced by other luminosity
sources.

\subsection{SED and Protostellar Model Global Parameters}\label{S:sed}

We integrated over a $40\arcsec\:\times\:50\arcsec$ region to find
total 31 and 37~$\mu$m fluxes of 527 and 922~Jy, respectively. No
significant emission was seen outside of this area in either band. We
constructed the SED using data from 2MASS (\citealt[]{Skrutskie06}),
GLIMPSE, Gemini-T-ReCS, IRAS-IGA (\citealt[]{Cao97}), Herschel-SPIRE
and JCMT-SCUBA (this paper), and IRAM 30m Telescope
(\citealt[]{Mooney95}). Total 2MASS and IRAC fluxes were integrated
from background-subtracted images in the same region as the FORCAST
bands. The 10 and 18~$\mu$m fluxes (3.06 and 46.87~Jy,
\citealt[]{Debuizer06}) were integrated with a 30$\arcsec$ aperture
(again, there are no significant sources in the immediate vicinity).
For IRAS bands, the beams are so large that the
$40\arcsec\:\times\:50\arcsec$ region does not contain all the flux
from this source, so we use larger apertures
($6.5\arcmin\:\times\:3\arcmin$) that encircle all the emission from
the source while avoiding fluxes from significant nearby sources.  We
obtained fluxes at 60 and 100~$\mu$m of $2.37\times10^3$ and
$4.35\times10^3$~Jy.  They may be considered as upper limits.  For
SPIRE and SCUBA images, we have carefully subtracted the backgrounds
with two or more component Gaussian fitting. The fluxes are 546, 420,
123, and 16~Jy at 250, 350, 450, and 850~$\mu$m respectively with
source sizes of $27\arcsec\:\times\:18\arcsec$ ,
$25\arcsec\:\times\:3.8\arcsec$ at 250 and 350~$\mu$m, and
$22\arcsec\:\times\:12\arcsec$ in SCUBA bands.  The SED is shown in
Fig.~(\ref{fig:sed}).

We use the SED to constrain the properties of the protostar, its core
and the surrounding clump by fitting with the RT models of ZT11 and
ZTM13. The models are developed self-consistently for a core embedded
in a high pressure clump forming a massive star via core accretion
(MT03), including a treatment of rotating infall, an active
accretion disk and a disk wind filling the outflow cavity. We have
further extended the ZTM13 models to also include emission from the
surrounding clump (see Fig.~\ref{fig:cartoon}). Since we have not yet
constructed a full model grid, we limit ourselves to achieving only a
moderately good fit by exploring the parameters manually and with a
quite coarse sampling of parameter space. We show four models that
have relatively good matches to the SED in Fig.~(\ref{fig:sed}) with 
their parameters listed in Tab.~(\ref{tab:models}).

Model 1 (Fig.~\ref{fig:cartoon}) contains a core of initial mass 
$M_c=240\:M_\odot$, and a protostar of mass $m_*=34\:M_\odot$ forming at its center. The
core is embedded in a clump with $\scl=1\:\gcm$, thus setting its
radius $R_c=0.11$~pc (MT03). We distribute the clump material between 1 and
$10\:R_c$ with a power-law density profile
(with index of -1.75, e.g., \citealt[]{Mueller02}) that joins the core
boundary smoothly and reaches $\scl\:\simeq\:1\:\gcm$ by $10\:R_c$
(counting two sides along a central sight line). Then the original
clump mass (assuming spherical symmetry, i.e. before some material is
swept-up by the outflow) is $4.8\times10^3\:M_\odot$.
The size of the disk is set by assuming a ratio of rotational to
gravitational energy of 2\%, yielding an outer radius of 940~AU. Its
mass is assumed to be 1/3 of the stellar mass. The accretion rate is
$5.1\times\:10^{-4}\:\myr$.  The total luminosity is
$2.2\times\:10^5\:L_\odot$ including protostar, accretion boundary
layer and disk. An outflow cavity, filled with a disk wind, including
dust along streamlines launched from the cooler disk outside of 7~AU,
is included. Following ZT11 and ZTM13, the cavity opening angle,
$\theta_{w,\mathrm{esc}}=51^\circ$, to yield a final star formation
efficiency from the core of 50\%. The viewing angle between the sight line
and the outflow axis is $\theta_\mathrm{view}=58^\circ$.  At
this viewing angle, if the received flux was assumed to be isotropic
then a bolometric luminosity of
$3.6\times\:10^4\:L_\odot$
would be implied.
After considering the foreground extinction as a free parameter to
better fit the NIR data ($A_V=8$~mag for Model 1) and applying an
appropriate aperture size ($\sim50\arcsec$), the inferred isotropic
bolometric luminosity based on the received flux becomes
$3.3\times\:10^4\:L_\odot$. 
Three other models with smaller values of $\scl$, $m_*$ and
$\theta_\mathrm{view}$ are also shown in Fig.~(\ref{fig:sed}) (see also
  Table~\ref{tab:models}).

Overall, good agreement is achieved between these model SEDs and the
observations.  Small discrepancies in the NIR and IRAC bands may be
due to patchy extinction and/or poorly-modeled polycyclic aromatic
hydrocarbon (PAH) features and/or thermal emission from transiently
heated small dust grains. At 60~$\mu$m the observed IRAS flux is lower
than the model (after convolving with the relatively broad filter
response of IRAS 60~$\mu$m band) by $\sim$25\%.  These results show
the SED fitting has a certain degeneracy in model parameter space,
especially when the opening angle is not well constrained (ZTM13 found
the outflow, especially the bright part, at 10 to 20~$\mu$m appears
narrower than the actual cavity, so the MIR morphology may not be a
reliable tracer of the total outflow opening angle).  Models with
narrower outflow cavities reprocess larger fractions of their
luminosities and thus have smaller bolometric flashlight effect
corrections for the considered viewing angles. Thus lower luminosity
protostars are able to reproduce the observed FIR peak of the SED.
Then, lower values of $\scl$ and $\theta_\mathrm{view}$ are
needed to match the MIR fluxes.  
Observations at 10~$\mu$m with higher spectral resolution will provide
strong diagnostics for testing these models.
However, even with a large
uncertainty in the outflow opening angle, the protostellar mass is
still estimated to be $\sim20\:-\:34\:M_\odot$ and
$\scl\sim0.4\:-\:1\:\gcm$, i.e., G35.2 is a massive protostar forming
from a massive core embedded in a high surface density clump.

\subsection{Resolved Intensity Profiles Along the Outflow Axis}\label{S:fluxdist}

The images of Model 1 are shown in Fig.~(\ref{fig:cartoon}) for reference.
The other three models selected from SED fitting 
show slightly narrower outflow cavities.
We compare the intensity
distribution along the outflow axis predicted by these
models with observations in Fig.~(\ref{fig:fluxdist})).
The axis direction of G35.2 is chosen via the
radio continuum and MIR morphology (dashed line in
Fig.~(\ref{fig:img}) with P.A. of $6^\circ$). The model profiles are
all convolved with the corresponding instrument beams. At each offset
from the center, we average over a perpendicular width of 2$\arcsec$
to estimate profiles for both model and data.
All model and observed profiles are normalized to the
average values over the near-facing regions that have significant
emission.
We also add constant background ambient intensities, which may be
either due to instrumental noise (i.e. in the T-ReCS data) or from
additional ambient interstellar material. Note we have not attempted to
use the lateral width of the cavities to constrain the models

We emphasize that a detailed model search to fit these profiles
separately from the SED has not been performed, but still Model 1
agrees very well with the observations at 37, 31, 18 and 10~$\mu$m,
producing the right peak positions and asymmetries of the two sides of
the outflow.  At 10~$\mu$m, Model 1 predicts higher extinction towards
the center. A possible explanation could be the strength (depth and
width) of the silicate absorption feature adopted in the dust model
used in the RT calculations (ZTM13).  In the IRAC bands, while a
reasonable fit is achieved for the near-facing outflow cavity, the
intensities of the far-facing side predicted by the model are too low
compared to those observed. Possible reasons for this include: (1)
non-uniform, possibly patchy, extinction from the clump or the adopted
uniform ``foreground'' extinction, that may actually be relatively
local to the source (e.g. a foreground extinction of $A_V=15$ lowers
3.6~$\mu$m flux by a factor of two); (2) enhanced emission from PAHs
and transiently-heated small dust grains in and around the outflow
cavity. Models 2, 3 and 4 also have good matches with the data in
the near-facing and far-facing wings at 37, 31, 18 and 10~$\mu$m, but
their profiles become too peaked towards the center due to the lower
extinction of their lower $\Sigma$ cores and clumps.

These results, especially the intensity profiles for Model 1 from
$\sim\:10\:-\:40\:\mu$m (for which the modeling has potentially fewer
problems associated with small grain and PAH emission), again support
the paradigm that G35.2 is a massive protostar forming from a high
$\Sigma$ core and clump.  The match of the SED and intensity profiles with
observations becomes less good when $\scl$ becomes lower. 
We estimate that $\scl$ should be no lower than $0.4\:\gcm$.  
The profiles also suggest that the accretion flow
and outflows are relatively well-ordered and symmetric.

\section{Discussions \& Conclusions}\label{S:conclusion}

SOFIA-FORCAST provides imaging with high dynamic range similar to
space-based instruments at a unique wavelength region $\sim$30 to
40~$\mu$m, where lower extinction allows us to search for the
predicted (ZT11, ZTM13) far-facing outflow cavity from a massive
protostar forming from a high surface density core. At longer
wavelengths the emission is predicted to become even more symmetric,
being dominated by cooler dust in the core/clump. Our SOFIA-FORCAST
observations at 31 and 37~$\mu$m did reveal emission from the
far-facing outflow cavity of G35.2, which was too faint to detect by
ground-based T-ReCs 11 and 18~$\mu$m observations.

We compiled the NIR to mm SED of G35.2. RT modeling of a massive
protostar forming from a massive core bounded by a high $\Sigma$ clump
gave good agreement with this SED for four models. Depending on the
outflow cavity opening angle (35$^\circ$ to 50$^\circ$), we found
$m_*\sim22-34\:M_\odot$, $L_\mathrm{bol}\sim(0.7-2.2)\times10^5\:L_\odot$
and $\scl\sim\:0.4-1\:\gcm$.
Model 1 also produced intensity profiles along the outflow axis
that fit the observations well at 10, 18, 31 and 37~$\mu$m, without
need for extensive fine tuning.  These results indicate G35.2 is a
massive protostar, forming from high surface density core and clump,
via relatively ordered, symmetric collapse and accretion. Powerful
bipolar outflows are being launched and have cleared wide-angle
cavities, that are also relatively symmetric.

A protostar with the luminosity estimated here
($\sim1\times10^5\:L_\odot$) is expected to drive a CO outflow with
momentum flux of $\sim0.1\:\myrkms$ (\citealt[]{Richer00}), which is
much larger than the observed value: even assuming that the larger
NE-SW CO outflow is also driven by this source, the total momentum
flux is still only $\sim0.003\:\myrkms$ (\citealt[]{Gibb03},
\citealt[]{Birks06}). This may indicate that the outflow from G35.2N,
being partly ionized as indicated by the observed radio continuum jet,
may be relatively deficient in CO emission. Note that the modest
misalignment of the larger CO outflow with the radio and MIR jet may
result from the interaction of the wider angle part of the outflow
with a core/clump that does have some moderate lateral asymmetries.

We conclude that, at least in the case of G35.2, massive star
formation appears to be well described by predictions of the Turbulent
Core Model (MT03), even with a highly idealized realization of the
model utilizing perfect axissymmetry and smooth density structures
(see \citealt[]{Cunningham11} for a numerical simulation of this model
including outflows).  Such ordered accretion and outflow is not
expected in models involving massive star formation via competitive
accretion or stellar collisions, nor indeed in cores that are highly
turbulent. This may indicate that magnetic fields are dynamically
important in setting the structure of the core.

\acknowledgements This work is based on observations made with the
NASA/DLR Stratospheric Observatory for Infrared Astronomy (SOFIA).
SOFIA is jointly operated by the Universities Space Research
Association, Inc. (USRA), under NASA contract NAS2-97001, and the
Deutsches SOFIA Institut (DSI) under DLR contract 50 OK 0901 to the
University of Stuttgart. YZ acknowledges support from a Graduate
School Fellowship from the Univ. of Florida.  JCT acknowledges support
from NSF CAREER grant AST-0645412 and NASA/USRA grant in support of
SOFIA Basic Science observations. CFM acknowledges support from NSF
grants AST-0908553 and AST-1211729 and NASA grant NNX09AK31G.

\end{document}